\documentclass[manuscript]{aastex}

\usepackage{emulateapj5}

\def\lesssim{\mathrel{\hbox{\rlap{\hbox{\lower4pt\hbox{$\sim$}}}\hbox{$<$}}}}

\def\gtrsim{\mathrel{\hbox{\rlap{\hbox{\lower4pt\hbox{$\sim$}}}\hbox{$>$}}}}

\begin{document}

\title{Discovery of new faint radio emission on 8$\arcdeg$ to 3$\arcmin$ scales in the Coma field, and some Galactic and extragalactic implications}












\author{P.P. Kronberg\altaffilmark{1,2}, R. Kothes\altaffilmark{3,4},
C.J. Salter\altaffilmark{5}, P. Perillat\altaffilmark{5}}

\altaffiltext{1}{IGPP, Los Alamos National Laboratory, Los Alamos NM USA}
\altaffiltext{2} {Department of Physics, University of Toronto,
 		60 St. George Street, Toronto M5S 1A7, Canada}
\altaffiltext{3}{National Research Council of Canada,
               Herzberg Institute of Astrophysics,
               Dominion Radio Astrophysical Observatory
               P.O. Box 248, Penticton, British Columbia, V2A 6K3, Canada}
\altaffiltext{4}{Department of Physics and Astronomy, University of Calgary,
              2500 University Drive N.W., Calgary, AB, Canada}
\altaffiltext{5}{Arecibo Observatory, HC3 Box 53995, Arecibo, PR
00612, USA}













\begin{abstract}

We present a deep, 8$\degr$ diameter, 0.4 GHz radio image using a first time
combination of the NAIC Arecibo 305-m telescope in Puerto Rico, and the wide-angle
interferometer at the Dominion Radio Astrophysical Observatory at
Penticton, Canada. Our observations are centered on the Coma Cluster of galaxies
in the ``Great Wall'' of galaxies near the North
Galactic Pole. The complementary nature of these two instruments
enables us to produce a distortion-free image that is sensitive
to radiation on scales from 8$\degr$ down to that of an individual
galaxy halo at the 100 Mpc distance of the Great Wall. Newly revealed patches
of distributed radio ``glow'' are seen
well above the detection limit. One prominent such area coincides with groupings of radio
galaxies near the Coma cluster, and indicates intergalactic IGM magnetic fields in the range
0.2 to 0.4 $\mu$G on scales of up to $\sim$ 4\,Mpc. Other patches of
diffuse emission, not previously explored at these high latitudes on arcminute scales, probably contain
Galactic ``cirrus''. A striking anticorrelation is found between low-level diffuse
radio glow and some regions of enhanced optical galaxy surface density, suggesting
that cosmological Large Scale Structure (LSS), normally defined by the
baryonic (or dark) matter density, is not {\it uniquely} traced by faint continuum radio glow.
Rather, intergalactic diffuse synchrotron radiation represents IGM magnetic and Cosmic ray energy density,
instead of matter density.  The diffuse, arcminute-level structures over a large region of
sky are potentially important pathfinders to CMB
foreground radiation on high multipole scales.

\end{abstract}

\section{Introduction}

The existence of magnetic fields in and near the periphery of clusters
of galaxies, and their implied $\mu$G-level strengths, leads
naturally to the question as to whether significant fields exist on
yet larger scales in the intergalactic medium (IGM), namely on the
scale of galaxy filaments in cosmological large scale structure
(LSS). If so, important objectives are (1) to investigate the
intergalactic magnetic field strength and structure, (2) to determine
the excitation state and metallicity of any associated warm-hot
intergalactic medium (WHIM) (e.g. Henry 2004, Nicastro et al.  2005),
(3) to understand the intergalactic Cosmic Ray spectrum (e.g.  Colgate
\& Li, 2004), and (4) to define the propagation environment of Ultra
High Energy Cosmic Rays (UHECR's) (e.g. Sigl et al. 2004). The
achievement of these goals is ultimately important for solving important puzzles
in high energy astrophysics, astro-particle physics, the origin of
cosmic magnetic fields, and the evolution of LSS. Evidence for diffuse low frequency ($\lesssim$ 30 MHz)
radiation in the Coma cluster vicinity at low resolution and low dynamic range has been presented
in earlier work (e.g. Roger, Bridle, \& Costain, 1973, Henning 1989). More sensitive and accurate
arcminute resolution images (e.g. Kim et al. 1989 (0.3 GHz), En\ss lin et al 1999 (74 MHz) have clarified the
distribution of diffuse radiation, hence intergalactic magnetic field strengths in
the near environs of the Coma cluster. The first attempt to probe magnetic field strengths on
larger local-universe scales via Faraday rotation is described by Xu et al. (2006).

Both the feedback of outflows driven by galactic black holes (BH) (e.g.
Furlanetto \& Loeb 2001, Kronberg et al. 2001, Gopal-Krishna \& Wiita
2001) and the gravitationally-driven evolution of large scale cosmic
filament structure (e.g. Ryu, Kang \& Biermann 1998, Dolag et al 2004)
affect the physics of the IGM. Therefore, to understand both galaxy
and LSS evolution over the redshift range of $0 < $ z $\la\, $15, it
is important to determine the strength and structure of
intergalactic magnetic fields on size scales $\sim$ 3 $-$ 100 times that of
galaxy clusters.

Faint synchrotron radiation is a sensitive
indicator of widespread magnetic fields. The detection of weak radio
emission from cosmic ray electrons gyrating in intergalactic magnetic
fields requires high sensitivity to {\it both} large and small scale
structures at frequencies low enough that the spectral density of
the synchrotron radiation is relatively high. While most radio
interferometers can provide sensitive images of small scale structure,
they inherently filter out structures broader than the angular size
corresponding to their shortest spacings. In contrast, the largest
single-dish antennas, with low noise receivers and vast collecting
areas, provide enhanced surface brightness sensitivity for the
detection of faint large scale emission.  However, these have
insufficient resolution to resolve blends of discrete radio sources. Motivated by
the topical  questions (1) - (4) above, we have combined the sensitivity of
a very large single reflector with a well calibrated, wide-field
interferometer to obtain a sensitive and precise image over
a range of angular scales from 3 $^{\arcmin}$ to those that overlap degree-scale
all-sky surveys.

\section{The observational approach}

To overcome the spatial frequency coverage problems mentioned in
$\S$1, we have used the unprecedented combination of the 305-m Arecibo
radio telescope and the wide-field interferometer telescope of the
Dominion Radio Astrophysical Observatory (DRAO) at Penticton,
observing at frequencies near 0.4 GHz. The resultant resolution is
comparable to that of a 1000-m single-dish telescope. Fortunately,
the area near the North Galactic Pole that we have imaged lies within
the limited declination range accessible to both telescopes. DRAO
interferometer data were also collected in spectral line mode near 1.4
GHz, though these are not discussed here.  The DRAO interferometer has
9-m diameter dishes and a maximum baseline length of 617 m. This
combination of the largest single antenna with the precision
imaging capabilities of the DRAO interferometer can produce
distortion-free images at 0.4 GHz on all angular scales from 3' to
8$^{\circ}$. The combination gives a large overlap in spatial
frequency, enabling an effective cross-calibration of the Arecibo
and DRAO data in $u$-$v$ space to optimize image reliability.

We chose to observe a section of the Great Wall
centered on the Coma cluster of galaxies (at z=0.02320, or
100$h_{70}^{-1}$ Mpc distance). The linear resolution at 100 Mpc is
$\sim $120 kpc at full resolution, close to a galaxy halo scale at the Great
Wall, and hence favorable for separating individual radio sources from
true diffuse radiation.

The DRAO interferometer, having only East-West spacings, has no
geometrical ``${\it w}$'' component. Consequently, in a single pointing
at 408 MHz it produces a distortion-free image over the entire
9.5$^{\circ}$ (to the 10\% response level) of the primary beam,
whose FWHM is 5.5$\arcdeg$. This, along with the large spatial
frequency overlap, makes it straightforward in principle to combine
the DRAO data with the raster-scanned Arecibo images to undertake
sensitive, high-precision, wide-area radiometry
in a single pointing. The result is a combination of image
fidelity and surface brightness sensitivity that is unprecedented for
this frequency and resolution range over a large field of sky.

\section{Observations and data processing}

\subsection{Observations with DRAO}

The DRAO antenna feeds are mounted at the prime focus and
simultaneously accept radiation at 1420~MHz in both circular
polarizations, and at 408~MHz in right-hand
polarization only. At the center of our maps the synthesized beam has
dimensions of $2\farcm8 \times 6\farcm7$ (R.A. $\times$ Dec.) at
408~MHz. A detailed description of the DRAO Synthesis Telescope can be
found in Landecker et al. (2000).

With the DRAO Synthesis Telescope, we observed two pointings towards
the Coma Cluster. These were centered at:
$\alpha$(J2000) = 12$^h$ 57$^m$ 55$^s$, $\delta$(J2000) = $+27\degr$ 39$\arcmin$ and
$\alpha$(J2000) = 12$^h$ 57$^m$ 55$^s$, $\delta$(J2000) = $+28\degr$ 09$\arcmin$, i.e.
the two field centers differed by $30\arcmin$ in Declination. This
served as an additional check against artifacts in low-level diffuse
structure, and provided a $\sqrt{2}$ improvement in overall signal to
noise.

A full synthesis at each pointing consisted of 12 individual 12~hr
observations. Each 12~hr run had a different configuration of the three
(of seven) movable antennas along a precision railway track.  Thus, the
measurements for the two pointings extended over a total of 24 12-hour days,
and took place in Spring 2003. Complex gains for the pointing
centers were calibrated by observing the unresolved sources 3C147 and
3C295. Their flux densities at 408~MHz are 48~Jy and 54~Jy, respectively. The rms
noise level on the image from the combined DRAO pointings was $\sim$
2.9~mJy/beam, which corresponds to $\sim$ 300~mK.

Phase and amplitude drifts over each 12~hr observing interval were
corrected via self-calibration. At this point, many strong compact
sources have remaining ring artifacts, and these are removed by a
procedure called MODCAL, which uses the CLEAN source components as
its input model. The image processing routines were developed
specifically for the DRAO Synthesis Telescope and are described in
Willis (1999).  Due to an interferometer's lack of zero spacing
information, negative bowls appear around extended sources. Residual
rings around compact sources located in these bowls are hidden because
the peak of the source is lower than its true value, and may even be
at a negative level.  To overcome this problem, extended and fine scale
structure were CLEANed separately at the end of the data processing.
Amplitude self calibration can produce slight flux density scale
errors in the individual observations. For this reason we used the
original unprocessed but properly calibrated image to re-calibrate the
final DRAO image using all point sources with flux densities
above 10~$\sigma$. The required corrections for each of the two
final images were less than 5\%.

\subsection{Observations with Arecibo}

The 305-m Arecibo telescope has been upgraded by the addition of
a 25-m dome containing secondary and tertiary reflectors. These
illuminate an effective
surface area of about 225-m diameter. The overall rms surface
accuracy, including the primary surface upgrade is $\sim $2 mm,
and the beam size is 10$\arcmin$ $\times$
12$\arcmin$ at 430 MHz.  The Arecibo observations were taken during
multi-night sessions in the springs of 2003 and 2004. The system
temperature was $\sim $ 55 K, of which $\sim $ 20 K comes from the sky
at the NGP. The receiver bandwidth was 20 MHz at a center frequency of
430 MHz. We recorded right- and left-hand  circular polarization
channels for each of four overlapping sub-fields. Each sub-field
was covered by two orhogonal (R.A./Dec.) sets of raster scans recorded
in spectral line mode.
After the removal of interfered frequency channels in both time and
frequency domains, and calibration, the orthogonal coverages were optimized using the
basket-weaving technique (Sieber et al. 1979).  Using reduction
techniques that were custom developed at Arecibo, the orthogonally scanned
sub-fields were then merged to form a single 8$^{\circ}$ by 8$^{\circ}$
image whose center coincided with that of the combined pair of DRAO
fields.  The Arecibo beam pattern was measured separately on the radio
bright galaxy 3C433, so
that the final Arecibo image could be beam-deconvolved.  Zenith
Angle-Azimuth variations of T$_{sys}$ at 430 MHz are well determined,
and were calibrated out.  The basketweaving and field combination
techniques were effective to the point where any residual striations
from raster scanning were so faint that they were a small fraction of
the final noise level and could no longer be seen visually in the
images.  The theoretical thermal noise level of our Arecibo image is
4.2 mK.  For absolute temperature calibration we used the the
408-MHz all-sky map of Haslam et al.(1982) with an adjustment,
described below, for the small difference in sky frequencies.

\subsection{Combination of the Data Sets}

The lower spatial frequency limit of an interferometer observation is
determined by the shortest antenna spacing present. For the DRAO
synthesis telescope this is about 13~m, which translates to about
$3\degr$ on the sky at 408~MHz. To achieve complete spatial frequency
coverage, the missing information on larger structures must be obtained
from our Arecibo 430-MHz observations, and for absolute
radiometric calibration we again used the 408-MHz all-sky map of
Haslam et al. (1982). Since the Arecibo map was at 430~MHz,
we had to extrapolate these data to the DRAO center frequency of
408~MHz.  This procedure was based on an inter-comparison of the flux
densities of compact sources above 1~Jy in the DRAO and Arecibo
maps. An average flux density ($S$) scaling factor of $1.05 \pm
 0.03$ was determined, corresponding to a spectral index of
$\alpha=-0.9$, defined by $S\propto \nu^\alpha$. The lowest and highest
individual ratios were 0.98 and 1.12. This is a reasonable
result for compact extragalactic radio sources.  After the appropriate
scaling, the Arecibo map was combined with the two individual DRAO
pointings separately, after suitable filtering in the Fourier
domain and applying the DRAO primary beam correction.

Before merging the single-dish and interferometer data, the
Arecibo image was converted into visibilities by Fourier-transforming,
and removing the Arecibo beam by dividing by the transform of the beam
pattern. The resulting image was multiplied by the DRAO primary beam
to create a short spacing map that complements the interferometer
data.  Then the two data sets were merged in the $u$-$v$ plane
using a normalized tapering function in the overlap area. At this
point the two resulting maps were primary beam corrected and then
co-averaged.

To confirm the flux density scaling, we used all compact sources above
200~mJy in the inner area of our final map. The flux densities were
compared with those published in Kim (1994). A linear fit gives a
gradient of $1.00\pm 0.02$ (see Fig. 1). The actual r.m.s. noise in the
final image at full resolution is about 360~mK, which translates to
3.2~mJy/beam. This includes ``confusion'' due to weak background
sources, residual scanning noise at Arecibo, residual fluctuations in
the Arecibo T$_{sys}$ calibration, an unknown Galactic disc plus halo
contribution, and any diffuse extragalactic emission.

\section{The images}

Fig. 2 shows a combined Arecibo + DRAO + Effelsberg image at the
full DRAO resolution of 2.8$\arcmin$ $\times $ 6.7$\arcmin$. It
excludes the outermost zone of the DRAO primary beam beyond an
8$\arcdeg$ diameter circle. This image clearly shows patches of
diffuse radiation. The patterns of diffuse emission are particularly
apparent after the subtraction of discrete sources. This is seen
in Fig. 3 which, like Fig. 2, is at the full resolution and has
complete {\it u-v} coverage. Figs. 4 and 5 present the image with
the discrete sources subtracted (as in Fig. 3) and smoothed to a resolution of
10$\arcmin \times$ 10$\arcmin$, close to the FWHM of the Arecibo
telescope. The images are discussed below.

\subsection{Checks on image reliability}

To further investigate the reliability and nature of the diffuse
patches, we have compared the independent DRAO pointings, and
examined the effects of removing the point sources. This tests
whether some of the diffuse patches could be a residual summation of
incompletely removed sidelobes of the many discrete sources.  We found
that discrete radio sources have no detectable effect on the level and
distribution of the diffuse structure.

Another image, not shown, was made using only the DRAO data (i.e.
without Effelsberg or Arecibo), and convolved to the 10$\arcmin$
resolution of the Arecibo telescope. Virtually all of the larger scale
features in Figs. 4 and 5 could be recognized, although with poorer
definition and signal to noise. Also, significantly, when the Arecibo
data were added with the optimal ($u$,$v$) overlap range (see below) no
new strong large scale features were introduced.

The possibility of Solar interference in the far-out sidelobes of the
DRAO array was also investigated. The Sun was above the horizon for
less than 20\% of the two DRAO observing runs. While risen, its
Hour Angle relative to the Coma field, $\Delta$(HA), was large, and
varied over the 2-month period of the DRAO observations. No
systematic Solar-related effects were identified.

Another conceivable cause of image artifacts could, in principle, arise
at the shortest spacings of the small DRAO dishes where direct
feed-to-feed electromagnetic interaction could occur. To test for this
effect, we re-imaged the DRAO data removing the shortest physical
separations in two steps ($<$ 25 m and $<$ 50 m).  The images were
unaffected, except for the expected small effects at the largest
image scales from removing genuine, low ($u$,$v$) components. We also
subtracted these images from the final maps (which removes all
small-scale structure) and inspected these for low-level
``striping'' that might have artificially produced diffuse patches.
We found no evidence of a spurious origin for any of the diffuse
features due to short-spacing interactions.

An important test of the Arecibo-DRAO combined image reliability is to
examine the effect of varying the upper and lower limits of the
overlap zone in ($u$,$v$) space between the two instruments. This is
important to investigate, since the 0.4 GHz Arecibo telescope beam,
although very stable at a given pointing, has systematic beam squint
and beam pattern variations over the range of [AZ-ZA] pointing
directions. Since our measured Arecibo beam averages out such
variations, a beam deconvolution will introduce systematic errors at
the larger $|u$,$v|$ values in the Fourier-transformed Arecibo
image.  By varying the $|u$,$v|$ overlap zones between the 305-m and
DRAO telescopes we were able to find an optimal maximum $u$-$v$ limit
for the Arecibo data below which no distortions are introduced in the
combined image. An additional, and largely independent, criterion is to
test for consistency with the smoothed ``DRAO-only'' image described
above. Our final Arecibo ``$|u$,$v|$ low-pass'' filter limit
was set at 75 m. This limit leaves considerable Arecibo-DRAO overlap,
and retains the Arecibo sensitivity to the most important large scales
in respect of diffuse emission.

\subsection{Nature and location of the diffuse patches}

In Figs. 3-5 we show the combined, full-resolution image from which we
have subtracted (CLEANed) the discrete sources, except for the diffuse
halo of the Coma cluster itself.

Several patches of diffuse emission can be seen both at the higher 3' DRAO
resolution in Figs. 1 and 2, and at the $\sim$ 10' resolution of the
Arecibo Telescope. In the absence of additional information these could be either galactic or extragalactic
in origin. They are seen for the first time at these faint levels and small scales
at high Galactic latitude. Many of them probably represent Galactic foreground emission,
while others might be regions of extragalactic emission at the
Great Wall, or beyond. In this section we discuss the features in some detail, in particular the
two particularly bright features, which are labelled A and B in Fig. 5.

At the outer periphery of our images, the DRAO interferometer's
image noise level deteriorates by a factor of $\sim$2 and the discrete source
subtraction there might lack the precision it has over the rest of the field.  Thus,
extended features close to the image periphery could contain some
imaging systematics.

The most prominent diffuse feature (A) near the Coma cluster's halo has a clear morphological connection
with the Coma cluster halo. This connection is seen at both resolutions, and is especially
evident in figs 2, 4, and 5.

The inner, brighter parts of this diffuse region immediately west of the Coma cluster have been 
independently
detected at 326 MHz by Kim et al. (1989), at 1.4 GHz by Deiss et al. (1997), and
at 74 MHz by En\ss lin et al. (1999). It also overlaps the diffuse, ``satellite'' X-ray source
in the 0.5 - 2.4 keV band (White, Briel and Henry, 1993). We explain in section 4.5 why the
X-ray surface brightness distribution is unlikely to correlate 1:1 with the radio emission of Region A.

Region A appears as a complex of enhanced radio emission that extends to a maximum of $\sim$ 2.1$^{\circ}$ (3.7 $h_{70}^{-1}$ Mpc)
west of the center of the Coma cluster's radio halo as defined by the 2.25K contour in Figure 4. Our most conservative estimate of this
limit is $\approx$ 1.4$^{\circ}$ (2.4 $h_{70}^{-1}$ Mpc).
At 10 $\arcmin$ resolution its surface brightness is consistently $>$ 2.25K over a large area (Fig. 4),
in contrast to other surrounding emission. Region A
is also partly superimposed on a ``landscape'' of other diffuse radiation at brightness
levels from $\la$ 1.4K  to $\sim$ 2K $\approx$ 7$\sigma$ (Figure 5). These are the 
extended regions to the north and south
of A, and to the southeast of the Coma cluster itself. The 2.25K contour in Figure 4 was chosen to
illustrate that region A contains a broad component of emission that is consistently elevated above this background,
even though its detailed
T$_{B}$ variations will be modulated by the underlying $\Delta$T fluctuations just
mentioned which are $\la$ 1K. Besides this broad component region A also contains complex stronger substructures having
T$_{B}$ up to 3K (15$\sigma$ above baseline confusion) and above. This can be seen in Fig. 5, and also the brightest
parts of the images in Figs 2 and 3 (both at 2.8$\arcmin \times$6.7$\arcmin$ resolution).
Both these brighter and patchier areas and the broader emission show a clear contiguity with
the Coma cluster's radio halo. Independently, region A also coincides approximately with an unusual grouping of
7 radio galaxies, {\it all} at redshifts near the Great Wall, and which
lie approximately within the boundaries of region A (Refer to Figs 2 and 5). For all of the above reasons we conclude that region A
and certainly its brighter components are most likely extragalactic, and associated with the Great Wall.
The uncertainty in region A's outer boundaries is due to possible superimposition with other large scale features mentioned above, which are possibly of Galactic origin.

A second diffuse feature, B, is clearly seen on each DRAO image. Its
extent and total flux density are even better defined when combined
with the Arecibo data. Feature B is the relatively bright diffuse
patch near $\alpha$= 12$^h$ 59$^m$, $\delta$ = +30.3$\degr$, centered
about 2.5$\arcdeg$ directly north of the Coma Cluster. It is best
seen in Figs. 4 and 5. Its dimensions ($\lesssim $ 1$\arcdeg$), and
relatively high surface brightness suggest that it too is
extragalactic, possibly a giant, relic radio galaxy.  Its integrated
flux density at 0.4 GHz is $\sim$1.5 Jy, and its peak surface
brightness is $\sim$ 4 K. It is illustrative to compare this surface
brightness with the rms sensitivity, $\Delta$T$_{rms}$ $\sim $0.14 K,
of the 1.4 GHz NVSS survey (Condon et al. 1998), currently a widely
used and benchmark radio source catalog. At 1.4 GHz, the peak T$_{B}$
of feature B, scaled with $\alpha$=--1 to 1.4 GHz would be 100 mK, or
0.33~mJy/beam in the NVSS survey. This is a factor of $\sim $7 below the
faintest reliably detected source in the NVSS survey (2.3 mJy at 1.4
GHz), and just below  $\Delta$T$_{rms}$ of the NVSS.

At the higher resolution of Figs. 2 and 3, the northern end of
feature B appears to contain a ``cluster'' of faint discrete
sources. The peak 0.4 GHz flux densities of these unresolved
sources are $\lesssim $3 mJy/beam at our $2\farcm8 \times 6\farcm7$
resolution.  Returning to our check for detectability in the NVSS,
their corresponding 1.4 GHz flux densities would be $\lesssim $ 1
mJy/beam (for $\alpha \sim$ -1), which is also below the NVSS
detection limit.

The comparisons above illustrate how even some of the brightest diffuse
emission features can escape detection in discrete radio source
catalogs. It further demonstrates the advantage of full $u$-$v$
coverage with high sensitivity at the lowest $uv$ spacings. Source B
has no obvious optical identification, which is not surprising in
view of its large angular size. It seems most likely that it
is near the Great Wall, or even beyond.  We think it
unlikely to be Galactic, on the basis of its relatively high surface
brightness this close to the NGP, and the small cluster of faint
discrete sources that appear projected against it (Fig.  2).
However, at this stage we cannot entirely rule out some Galactic
contribution.

Many other apparently diffuse features appear unrelated
to regions A and B. They could be Galactic foregrounds seen for the first
time at this sensitivity and resolution, and/or combinations of extragalactic
diffuse emission and aggregates of faint, unresolved discrete sources.

\subsection{Comparison with nearby galaxy distributions}

Fig. 4 shows the combined DRAO-Arecibo image after removal of the
discrete sources and convolved to a 10$\arcmin \times $10$\arcmin$
beam. Here, and in Fig. 5 we have also removed a 22K $-$ 16K linear plane
approximation to the very broad emission structure derived from the
Haslam et al. (1982) 408-MHz survey.

To compare the diffuse
patches with optically visible galaxies, we show an overlay of galaxies
in the CfA2 catalog, color coded by radial velocity slice between 4,000
and 10,000 km/s (Huchra et al. 1990, de Lapparent et al. 1991).  The
radio glow in region A coincides with a galaxy overdensity zone to
the west of the Coma Cluster, although the correspondence is not exact.
This lack of detailed correspondence between diffuse radio emission and
Great Wall galaxies appears to be general within this field. This
statement can be further tested with a search for other galaxy
clusters. Our field is large enough to include two additional clusters,
having $v$ = 38,000  and 70,000 km/s, respectively, in an extension
of the CfA survey (Huchra et al. 2004).  These are approximately
superimposed, lying within 03$^{h}$ 06$^{m}$ $< \alpha <$ 03$^{h}$
10$^{m}$, 29$^{\circ}$ $< \delta <$ 30$^{\circ}$(inset in Fig. 4).
We find this region of high galaxy column density to have no
detectable diffuse radio emission above $\sim $ 200 mK. This shows that
dense groupings of galaxies are not {\it uniquely} associated with
regions of radio glow, although clearly this does happen in some
galaxy clusters and groups that have anomalously strong radio halos, such as
the Coma cluster.

\subsection{Comparison of low-level glow with discrete radio source locations}

Fig. 5 shows the same radio emission as Fig. 4, again at
10$\arcmin$ resolution, on which we have superimposed the strongest
discrete radio sources from Fig. 2.  Some groupings of discrete radio
sources appear to correspond with areas of radio glow, and especially
with region A. We recall that radio glow due to image
spillover from these discrete sources has been ruled out.

Conversely, the intensity minima of our image, discussed below, show
a tendency to not contain strong discrete radio sources. This apparent
trend is interesting, but needs statistical confirmation from
additional, similarly imaged, fields. This we are now
undertaking.  Discrete extragalactic radio sources are galaxies
containing AGNs which, over a Hubble time, are expected to ``feed
back'' energy into the IGM in the form of ejected cosmic rays (CRs)
and magnetic fields.  Global values for the intergalactic magnetic
field strengths in galaxy filaments of LSS due to AGN feedback have been
estimated to be of order 10$^{-7}$ G (Kronberg et al. 2001), and
the global energy released in magnetized CR electron plasma is
comparable to that released via quasar photons (e.g.  Choksi and
Turner, 1992). Generally, the discrete radio sources appear to be
projected against the zones of radio glow. This trend, most apparent
in Region A -- see below, suggests that we may for the first time be seeing direct evidence in
support of the BH$-$IGM feedback scenario in which the gravitational
infall energy of central BH's is injected into the surrounding IGM via
radio jets and lobes.  If so, over time the CR electrons in these zones
will eventually cease to be visible at cm wavelengths, but will retain
much of their magnetic energy, and possibly their CR proton energy, over
much of a Hubble time.

\subsection{Physical characteristics of region A}

In this section, for the reasons discussed in section 4.2, we
proceed on the assumption that region A is a truly extragalactic complex at
the approximate distance of the Great Wall, and not a foreground Galactic feature.
Because it is also superimposed on other, we presume unrelated, extended features
its outer boundaries cannot be precisely delineated, but this level of uncertainty
does not significantly affect our discussion which follows.

The observed 0.4 GHz surface brightness leads to approximate estimates
of intergalactic magnetic field strengths and total energy in region A
assuming equipartition of energy between relativistic particles and magnetic fields.
Adopting a parameter-space range that allows for different underlying
background contributions to T$_{B}$ in region A (giving
total 408 MHz flux density extremes of $\sim$ 0.8 and 3 Jy), a volume filling factor
of 0.1 and relativisic proton/electron energy ratios ($k$) ranging from 1 to 100, we
derive magnetic field strengths in the range of 0.2 to 0.4 $\mu $G in this intergalactic region.
The corresponding total energy is $\sim$ 10$^{59}$ to 10$^{60}$ ergs.

The radiating electrons do not appear to come from a single radio galaxy,
and it is likely that the radio glow in region A comes at least partly
from the aggregate activity, present and past, of radio galaxies near
the redshift of the Coma cluster. To illustrate this, we identify a
selection of co-located radio galaxies in Fig. 5, with their radial
velocities listed in the figure caption. These are taken from the
radio-optical identification list in Table 8 of Kim et al. (1994), to
which we have added 3C277.3, (No. 8, at the unrelated $v$ = 25583 km
s$^{-1}$), as a convenient reference source.

An additional energization could come from heating caused by
gravitational infall as suggested by the distributed X-ray emission
in this general area. An analysis of ROSAT all-sky survey PSPC data
around Coma (Briel et al. 1992) shows diffuse X-ray emission
covering part of region A. In projection, the diffuse X-ray
emission contours presented by Briel et al. ``fill'' part, but not
all, of the radio glow region A.  A visual
intercomparison of our Fig. 5 with Fig. 1 of Briel et al. is
facilitated for the reader by the common location of source No. 9
(NGC4789) in Fig. 5 with its location in the X-ray image. It lies at
the limit of region A and well beyond the outer X-ray contour
of the diffuse ``Coma-companion'' X-ray source. However, Briel
et al.  report that ROSAT PSPC data reveal further, faint, and
presumably unquantifiable, X-ray glow out to about 1.6$\arcdeg$ from
the cluster center. This finding has been confirmed by Finoguenov et al.
(2003) from XMM data.

The diffuse X-ray emission west of Coma appears to be mostly thermal (Briel et
al. 1992, Finoguenov et al. 2003). However, the CR electron energy
losses from region A will occur mostly by inverse Compton (IC)
scattering of CMB photons, as we demonstrate below. This raises the
question as to whether some (perhaps small) fraction of the X-ray
luminosity seen by Briel et al. (1992) could be IC emission from region
A.  The electron energies ($\gamma_{e}$ = $E_e$/mc$^{2}$) in
equipartition with a 0.3 $\mu$G magnetic field range from $\gamma_{e}
\sim$ 10$^4$ to 10$^5$, so that the up-scattered CMB photons are
broadly in the range of 0.3 MeV. This would make them virtually undetectable
at the $<$ 10 keV energies where the ROSAT, Chandra and XMM satellites
are most sensitive, though some keV-level photons could be produced if
$\gamma_{e}$ $\lesssim$ 10$^3$. This is consistent with the finding that at least most
of the 0.5 - 2 keV X-rays in this region are mostly thermal in origin,
and not more than a modest fraction can come from the CR electrons that
we detect in region A.  It suggests that future radio$-$X-ray
comparisons to higher energy X- and $\gamma$-ray bands will reveal
much more about the physics of this interesting and relatively
energetic IGM zone. It also confirms, as we find and would expect, that the thermal Coma satellite X-ray
emission does not mirror the details of the radio brightness structure of region A.

The requisite instrumental sensitivity in these
higher energy photon bands is unfortunately not yet available.  In future,
very precise determinations of the thermal
fraction of 0.5 - 2 keV X-rays from region A could in principle be
used to test equipartition, by limiting the low-$\gamma$ end of the CR electron
spectrum.

The energization associated with the thermal X-ray emission is thought
to come from the infall of at least two subgroups toward the Coma
cluster. These are associated with NGC 4839 and 4911 (Colless and
Dunn, 1996). Another large scale gravitational energy source could be
from infalling WHIM gas of neighboring cosmic filaments. The latter
are broadly consistent with the simulations of Miniati et al. (2000), and
are also concordant with the hot X-ray gas seen around the Coma cluster
by Finoguenov et al. (2003). All of this suggests that the CR heating
of the synchrotron-emitting electrons in Region A may have two sources;
(1) that from AGN driven outflow discussed above that is ultimately
provided by gravitational infall energy into the galaxy nuclei,
and (2) from large scale gravitational infall
that is implied from the velocity fields and perhaps the hot X-ray emitting gas.

It is interesting to compare estimated CR diffusion times with the
electron energy loss times (synchrotron and IC) at 0.4
GHz. If the CR's have diffused from a multiplicity of AGN in the area
as speculated above, the CR diffusion times, $\tau_{D}$, for an average
distance to the nearest AGN inside the $\sim $2 - 3.8 Mpc dimensions of
region A is the typical transport distance, $L$ divided by the
diffusion speed $V{_D}$. Normalizing to $V{_D}$ = 1000 km/s, comparable
to an IGM sound speed, and $L$ $\sim $ 1.8 Mpc, $\approx$ half the size
of region A, we get

\begin{equation}
\tau _D  \approx 1.7 \times 10^{9} \left( {\frac{L}{{1.8\,{\rm{Mpc}}}}} \right)\left( {\frac{{1000\,{\rm{km/s}}}}{{V_D }}} \right){\rm{yr}}
\end{equation}

This can be compared with the electron radiative loss times,
$\tau_{R}$, which in this system are dominated by IC
scattering against CMB photons:

\begin{equation}
\tau _R  \approx \frac{{2.8 \times 10^{8} B_{0.3\mu G}^{1/2} }}{{\left\{ {B_{cmb}^{2} (1 + z)^{4}  + (0.3B_{0.3\mu G})^{2} } \right\}\left\{ {\frac{{\upsilon (1 + z)}}{{0.4GHz}}} \right\}^{1/2} }}{\rm{~yr}}
\end{equation}

where $B_{cmb}$ $\simeq $ 3.3~$\mu$G is the magnetic field that is
equivalent in energy density to the CMB radiation. For the parameters
adopted above, $\tau_{R}$ is roughly an order of magnitude less than
$\tau_{D}$.  Unless relativistic particle diffusion times from their
host galaxies are far above 1000 km/s, i.e., well above the IGM sound
and magnetosonic speeds, this suggests that {\it in situ} acceleration
of CR electrons exist in the rarefied magnetoplasma of
intergalactic space {\it outside} of individual radio source lobes and
galaxy clusters. Evidence for distributed particle acceleration has
previously been deduced within the largest giant radio galaxies (e.g.
Kronberg et al. 2004) which can have comparable dimensions to region A.  In some
cases, these have been imaged near 10 GHz, where $\tau_{R}$ is
significantly smaller. At our imaging frequency of 0.4 GHz, this
apparent contradiction could be removed if, for example, $V{_D}$, poorly
known at this stage, were 10,000 km/s (3\% of $c$).

This comparison illustrates the importance of extending comparable
observations to $\ga $ 10 GHz using a next-generation
radio telescope such as the proposed ``Square Kilometer Array (SKA)'', and at
larger redshifts where radiative (IC) lifetimes are significantly shorter.
Assuming the extragalactic dimensions of Region A, our observations place
interesting constraints on the combination of
$V{_D}$ and $\tau_{R}$. Future, similar observations at higher frequencies, can
provide tighter constraints on
$V{_D}$ for an intergalactic CR magnetoplasma, since $B^{2}_{cmb}$ is
known exactly, and $B_{equipartition}$ can be reasonably well
determined. This may constitute one of the few possibilities for
{\it experimentally} estimating CR diffusion/transport speeds in
extragalactic space.

\subsection{Comparisons with CMB fluctuations at $\approx $ 30 GHz}

It is of interest to extrapolate the surface
brightness fluctuations in Figs. 3 \& 5 to 30 GHz, at which frequency CMB
fluctuations are imaged with similar angular scales. Assuming a constant
radio spectral index of $\alpha$ = -1, fluctuations
of our diffuse radiation at 0.4 GHz, whether Galactic or
extragalactic in origin, translate to $\Delta T$ $\approx $ 6.2
10$^{-6}$ K, at 32 GHz, or $\Delta T$/$T_{cmb}$ $\approx $  2.3
10$^{-6}$.  If most of the $\Delta T$ fluctuations in our images are
extragalactic, we would not expect to see background synchrotron
radiation fluctuations at $\approx $30 GHz on these scales unless there
is some space-filling re-acceleration mechanism on supragalactic
scales.  This adds further interest to the question of IGM {\it in
situ} acceleration raised in the previous section. Alternatively, if
some of the arcminute-scale fluctuations are Galactic foreground, there
is a greater possibility that a CMB foreground $\Delta T$/$T$ effect
could persist into the GHz range.  Alternatively stated, to how high in
frequency can the fluctuations seen in our maps be detected? This
question again stresses the importance of future maps at frequencies
$\gg$ 1 GHz to complement the capabilities of future low frequency
telescopes operating below $\sim$ 100 MHz. It also highlights the
importance of searching for, and identifying, foregrounds to CMB
fluctuations. To achieve this, one would compare images such as those
here near 400 MHz with others at successively higher frequencies in
the 1 - 30 GHz range. In any case, the low-level fluctuations in the
0.4 GHz diffuse emission reported here raise the possibility that
these might also be visible at substantially
higher radio frequencies, even if not to 30 GHz. Evidence
for such fluctuations in our companion DRAO 1.4 GHz observations is presently being analysed.

Such observational capabilities at both low and high radio frequencies
should be possible with an appropriate version of the proposed SKA
and the next generation of low frequency arrays. The possibility of
distributed acceleration in the IGM also points to the
need for a better understanding of plasma physics processes, such as
large scale magnetic reconnection and similar field-to-particle energy
conversion processes that might energize relativistic electrons over
large regions of space. If some of the diffuse features in Figs. 2 - 5
are Galactic foregrounds as seems likely the consequently, much smaller, physical
scales will permit relativistic electrons to radiate locally at much
higher frequencies for certain space-filling electron acceleration
scenarios. In that case, some CMB foreground radiation at frequencies
around $\sim$ 30 GHz is more likely.

The small scale features that we detect down to $\sim $4 $\arcmin$ scales,
and especially their polarized component (which we have not yet measured),
are potentially important contributions to CMB foreground at
high multipole scales ($l$ of order $\sim$ 1000) at $\sim $ 30 GHz. The
ultimate measurement of Stokes parameters I, Q, and U for any CMB foreground
is important for isolating vector (B) modes in the CMB that are potentially
important for tracing primordial, pre-recombination magnetic fields (e.g. Kosowsky \& Loeb, 1996).

\subsection{Minimum temperature zones}

Our images also show newly defined, ``cold'' areas at
both 2.8$\arcmin$ $\times $ 6.7$\arcmin$, and 10$\arcmin$ $\times $
10$\arcmin$ resolution, which contain relatively few discrete radio
sources. These are centered around ($\alpha$= 12$^h$ 53$^m$, $\delta$ =
+30$\degr$, No. 1), ($\alpha$= 12$^h$ 55$^m$, $\delta$ = +26.5$\degr$,
No. 2), and ($\alpha$= 13$^h$ 05$^m$, $\delta$ = +28.5$\degr$, No. 3).
Each $T_{B}$-minimum zone is $\sim$ 1$\arcdeg$ or more in size. For
regions (1), (2) and (3), the mean sky temperature levels are
$\sim$ 200, +50, and -200~mK respectively, after subtraction of the
22K $-$ 16K linear plane of ``baseline'' emission over our 8$\degr$
field that was independently derived from the Haslam et al. all sky survey (Sec. 4.3).
These residual ``floor'' temperatures are now comparable to
each other even though they
differ by 2 - 4$\arcdeg$ in location. This provides additional confirmation
that these regions contain true extragalactic base temperature levels. As we
show below, these minimum sky temperature levels also agree with
predicted confusion levels from discrete extragalactic source counts at 0.4 GHz.

At the resolution of 10$\arcmin$ $\times $ 10$\arcmin$ (Figs. 4 and
5), the rms fluctuations in these colder zones, $\sigma_{cold}$, are
about 0.2~K, less than 1 \% of the broad Galactic emission that we
have subtracted. We stress here that small
``negative'' temperatures in these zones do not represent a negative
instrumental brightness, rather they are consistent with noise, and expected,
being only of order $\sigma_{cold}$ below the baseline level of
$T_{B}$.

We now compare the rms $\Delta T$ levels in zones
(1) - (3) with the predictions of extragalactic point source confusion
at 0.4 GHz. The rms confusion level at a frequency $f_{GHz}$ for a
FWHM of $\theta_1^{\arcmin} \times \theta_2^{\arcmin}$ can be
approximated to,

\begin{equation}
\sigma_{conf}  \simeq 0.17{\rm{mJy}} \times \theta_1^{\arcmin}  \times \theta_2^{\arcmin}  \times f_{GHz}^{ - 0.7}
\end{equation}

\noindent
(Condon, 1987). For a 10$\arcmin$ $\times $ 10$\arcmin$
beam and our average frequency of 0.42~GHz, equation (3) translates to
$\sigma_{conf}$ = 0.6~K. After allowing for the subtraction of all
recognizable point sources from the full resolution image of Fig. 2
before smoothing to the 10$\arcmin$ $\times $ 10$\arcmin$ resolution
of Figs. 4 and 5, an rms confusion on the resultant image of
$\sim$0.25~K is expected. This is close to the values measured.

\section{Summary and Conclusions}

We have described the first attempt to image a large field with a
combination of the Arecibo radio telescope and the DRAO
Interferometer, which have significant aperture overlap between the
large single reflector and the interferometer. At resolutions as
fine as 3$\arcmin$, we can trace faint radio glow all the way to
degree scales in the region of the North Galactic Pole. The radio glow
to the west of the Coma cluster, region A, is likely to be, as we assume,
at the distance of the Great Wall, in which case it traces CR electrons and magnetic
fields out to the largest distance yet seen from a galaxy cluster.
Limits to the synchrotron and IC radiative lifetimes at
0.4 GHz combined with the dimensions of region A suggest that the CR
electrons are being accelerated, or re-accelerated, in intergalactic
space. Such evidence from region A parallels that from the comparably
sized lobes of individual giant radio galaxies (e.g. Kronberg et al.
2004). We have discovered a second, mostly diffuse feature (B) which,
even though relatively bright in our images, has not
been previously detected.

The origin of other degree-scale features is uncertain or unknown.
They could be either faint, small scale, previously unmapped
Galactic foreground features at the distance of the Great Wall, at
greater intergalactic distances including blends of fainter discrete
sources or, most likely, some combination of these. It appears that at least part of
region A is at the distance of the Coma supercluster, implying
that such glow regions should exist elsewhere in intergalactic space.
A possible relationship to the ``WHIM'' awaits better specification of hot
thermal gas in the IGM, e.g. via future EUV and X-ray images and high
excitation X-ray line location.

A preliminary intercomparison of the radio features
with both radio-loud and radio-quiet galaxies suggests that the
diffuse radio glow is probing LSS in a different
way to that defined by the large scale optical surveys of galaxies.

Intergalactic synchrotron emission presents an
interesting opportunity to probe plasma conditions in the IGM, and
especially intergalactic CR diffusion speeds. Our results suggest that
reconnection, or some similar CR acceleration mechanism, is occurring
to re-accelerate CR electrons in intergalactic space, as appears to
happen within comparably sized giant radio galaxy lobes.

\section*{Acknowledgments}

We thank John Huchra for his assistance with the CfA catalog
data, and Raul Cunha at the University of Toronto for his help with the
figures. We acknowledge helpful discussions with T. Landecker, R.S. Roger,
C.R. Purton, M. Wolleben, and A.G. Willis. This research was supported by the Natural
Sciences and Engineering Research Council of Canada (PPK, RK), the U.S.
Department of Energy through the LDRD program at LANL (PPK), and the National Research Council of
Canada. The Dominion Radio Astrophysical Observatory is a National
Facility operated by the National Research Council of Canada. The
Arecibo Observatory is part of the National Astronomy and Ionosphere
Center, which is operated by Cornell University under a cooperative
agreement with the National Science Foundation.


\begin{figure*}
\centering
\includegraphics[bb = 35 105 562 590,width=12cm,clip]{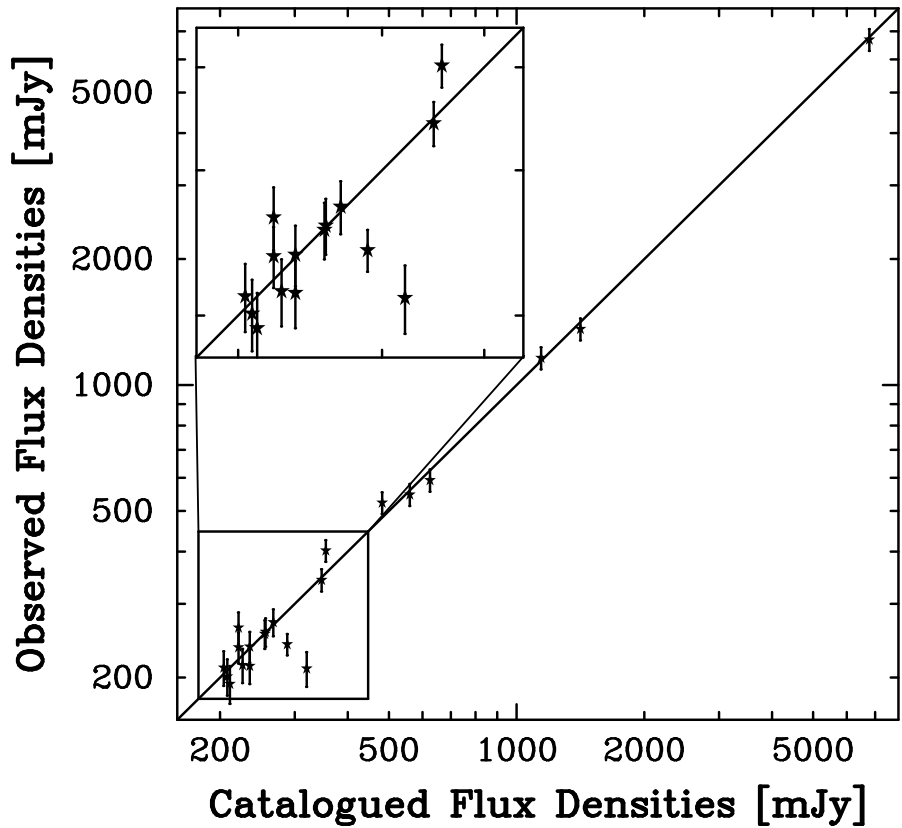}
\caption{A comparison of flux densities for compact sources brighter than 200~mJy
  at 408~MHz between the present observations and the values published by Kim (1994).}
\label{calib}
\end{figure*}

\begin{figure*}
\centering
\includegraphics[bb = 10 10 460 430,width=120mm,clip]{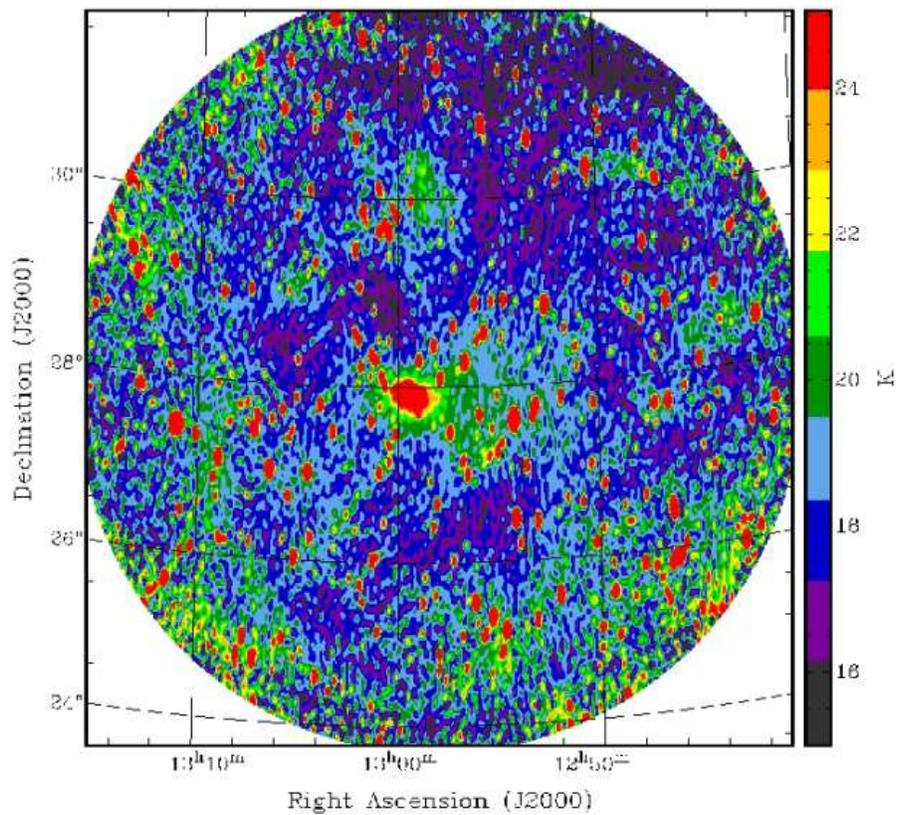}
\caption{Radio continuum image at 408 MHz of an 8$\degr$ diameter field within the Great Wall ``filament'' of galaxies
centered near the Coma cluster. The half-power beamwidth is 2.8$\arcmin$  $\times$ 6.7$\arcmin$ , elongated N-S.
The image contains discrete radio sources which are distinguishable at this resolution from patches of diffuse radio glow.}
\label{tpfin}
\end{figure*}

\begin{figure*}
\centering
\includegraphics[bb = 10 10 410 390,width=140mm,clip]{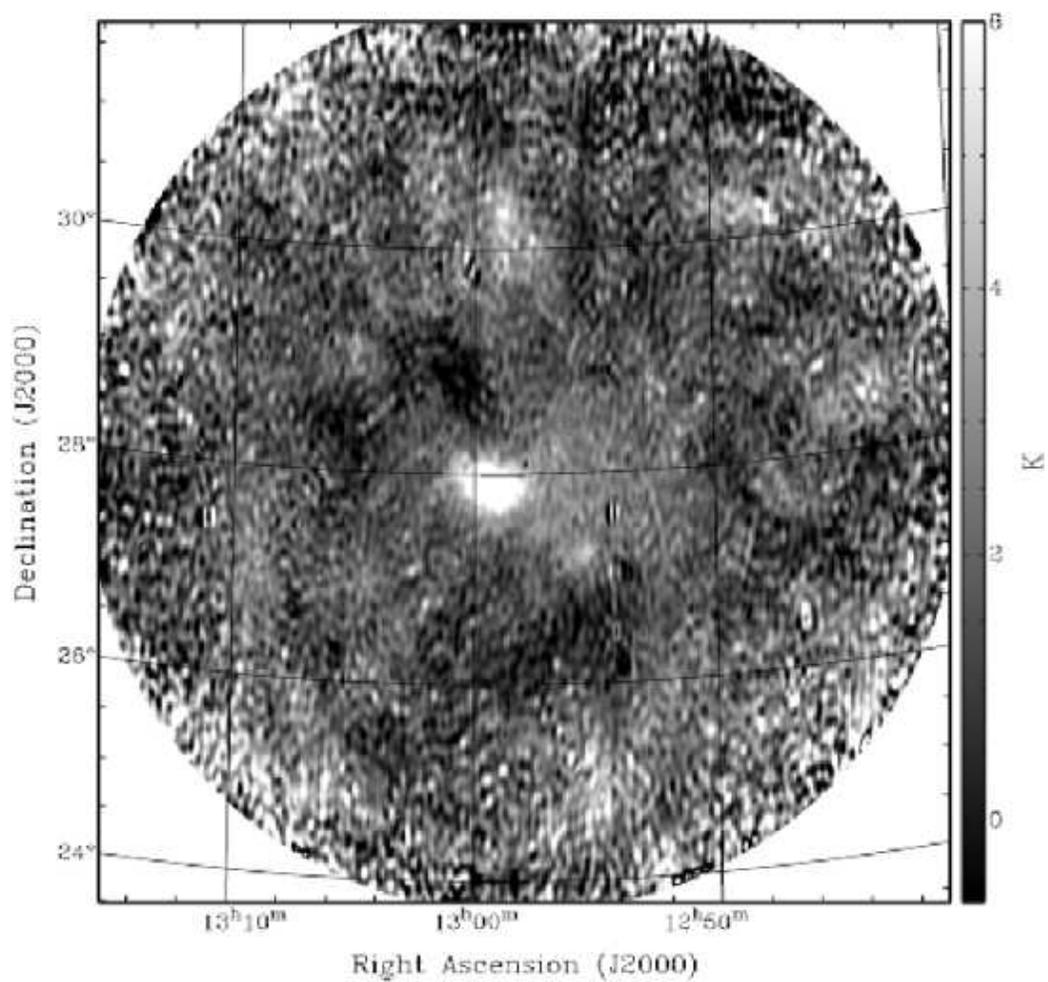}
\caption{The same image and resolution as in Fig. 2, but with the discrete sources and a smooth (CMB $+$ Galactic) background subtracted.}
\label{nops}
\end{figure*}

\begin{figure*}
\centering
\includegraphics[bb = 55 175 555 620,width=14cm,clip]{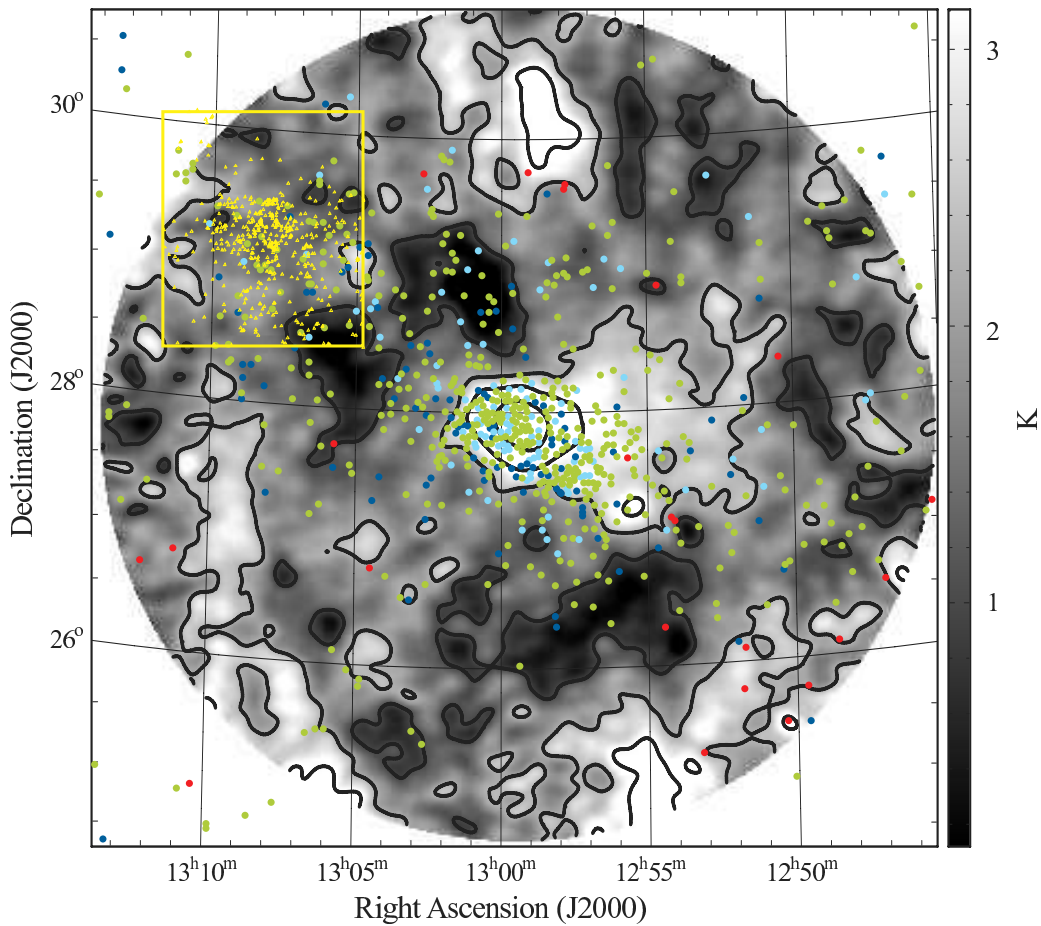}
\caption{The radio continuum image of Fig. 3 in grey scale, convolved to
a circular beam of half power width 10$\arcmin$. Contours levels are at 1.0, 2.25, 3.5, and 10 K above the mean level of the
three (comparable) coldest zones after subtraction of
the distributed extragalactic component, the CMB, and and a linear plane Galactic foreground. ``Negative'' values below
this mean are only of order $\sigma$ about this mean (see text). Overlaid are the positions
of Coma supercluster galaxies from the CfA catalog from Huchra et al. (2004). The galaxy redshift coding is red for velocities
below 4000 km/s, blue between 4000 and 6000 km/s, green between 6000 and 8000 km/s, and light blue between 8000 and 10,000
km/s. The yellow-coded points in the upper left inset box have radial velocities from 30,000 to 100,000 km/s, beyond the Great Wall, and are part of a CfA catalog extension (Huchra et al. 2004). The majority are a superposition of two
concentrations around radial velocities of 38,000, and 70,000 km/s (see text).}
\label{redshift}
\end{figure*}

\begin{figure*}
\centering
\includegraphics[bb = 10 10 435 435,width=14cm,clip]{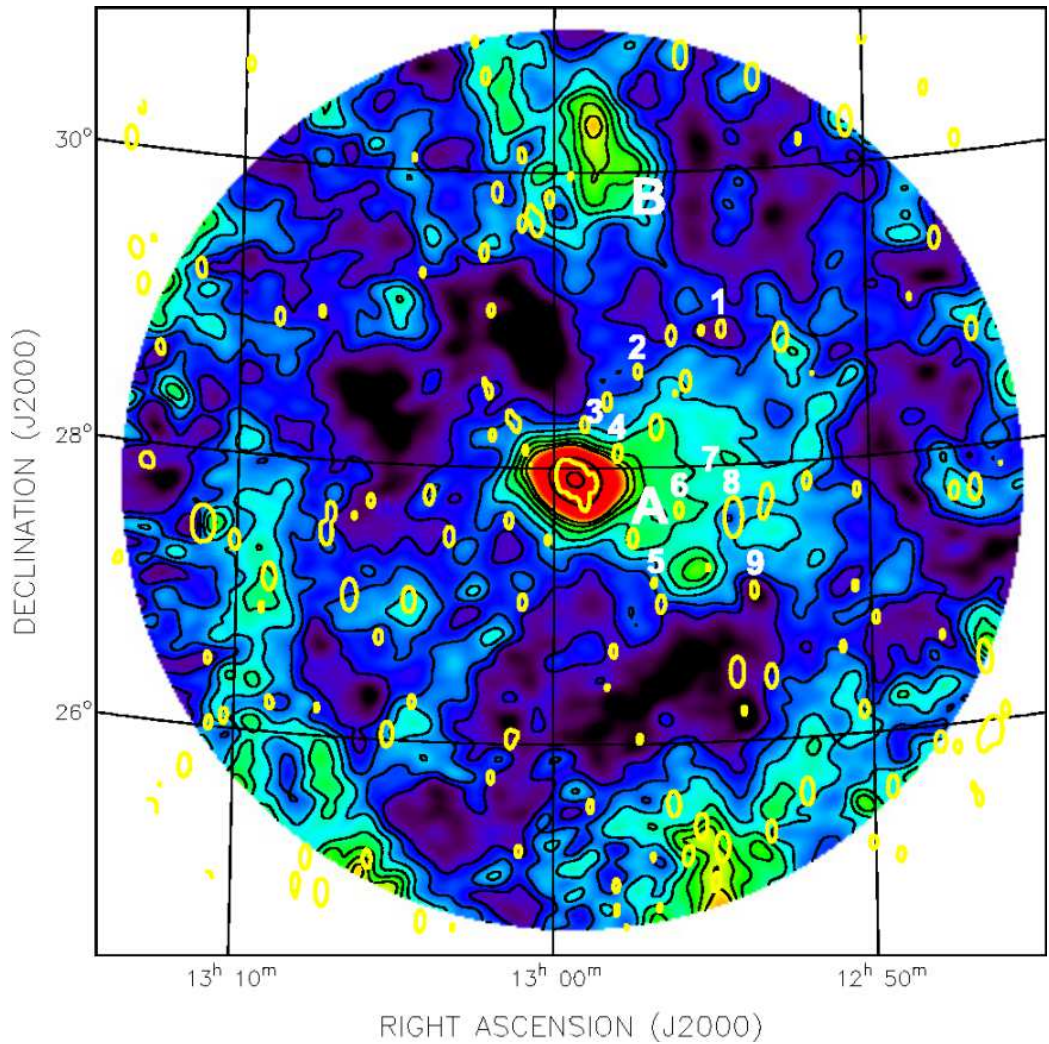}
\caption{The radio continuum image convolved to 10$\arcmin$
as in Fig. 4, now showing the T$_{B}$ levels in more detail, and overlaid with the brightest compact structures from Fig.2. The black contours are shown at  1.4, 1.9, 2.4, 2.9, 3.4, 3.9, 4.4, 10, 40 K above the same reference level as for Fig. 4. Yellow contours are overlaid to indicate the brighter discrete sources in the original map of Fig. 2. The numbered radio galaxies and their radial velocities are selected from Table 8 of Kim et al. (1994). They are (name, radial velocity in km s$^{-1}$): 1. N4793, 2465; 2. Zw160-58, 7649; 3. N4848, 7248; 4. anonE/S0, 8160; 5. N4827, 7356; 6. Zw160-20, 4900; 7. Zw160-15, 7470; 8. 3C277.3, 25583; 9. N4789, 8300. (Refer to discussion in text.)}
\label{gal}
\end{figure*}

\end{document}